%% file: ANA-OTRC-2025-01-PAPER.tex
\documentclass[cernpreprint, texlive=2025, atlasdraft=false, UKenglish, texmf, orcidlogo]{atlasdoc}
\usepackage[subcaption,full]{atlaspackage}
\usepackage{atlasbiblatex}

\usepackage{atlasphysics}
\graphicspath{{logos/}{figures/}}

\AtlasTitle{Breaking barriers: the impact of ATLAS Virtual Visits in science communication}

\AtlasAbstract{%
The ATLAS Collaboration at CERN’s Large Hadron Collider is at the forefront of particle physics research and is equally committed to bridging the gap between cutting-edge science and the wider public. Since 2010, the ATLAS Virtual Visits programme has provided live, interactive tours of the ATLAS detector and control room to global audiences in their language, without the need to travel. The programme has grown significantly since its inception, as demonstrated by quantitative data collected since January 2019 and case studies of large-scale implementations in Brazil and Greece. The impact on individual participants is also discussed.
}

\AtlasRefCode{OTRC-2025-01}

\PreprintIdNumber{CERN-EP-2026-013}

\AtlasJournalRef{Eur. Phys. J. Plus 141 (2026) 874}
\AtlasDOI{10.1140/epjp/s13360-026-07771-w}

\hypersetup{pdftitle={ATLAS document},pdfauthor={The ATLAS Collaboration}}
\begin{document}

\maketitle

\tableofcontents

\section{Introduction}
%


The Large Hadron Collider (LHC)~\cite{Evans:2008zzb} at CERN, near Geneva, Switzerland, is a particle accelerator with a circumference of 27 kilometres, making it the largest constructed to date. It accelerates protons and heavy ions at velocities approaching the speed of light, colliding them at the highest energy ever achieved by an accelerator. The primary objective of the LHC is to investigate the fundamental constituents of matter and to explain the physical laws that govern their interactions, addressing outstanding questions such as the origin of mass and the difference between matter and antimatter. The ATLAS detector~\cite{PERF-2007-01} is one of the four major detectors at the LHC, designed to measure the products of high-energy particle collisions. Standing 25 meters high and 46 meters long, in a cavern 90 meters underground, ATLAS is the largest among the LHC experiments and was constructed as a general-purpose detector to study a broad range of physical processes. Since the start of the programme in 2008, the ATLAS Experiment has collected extensive datasets, resulting in over 1,400 scientific publications. As one of the largest scientific collaborations in the world, ATLAS comprises approximately 6,000 physicists, engineers, technicians, students, and administrators from over 200 institutes worldwide, representing more than 40 countries and close to 100 nationalities~\cite{Collaboration_FactSheet}.

Particle physics has proven effective in engaging a wide audience, including students, educators, and science enthusiasts.
Its accomplishments leave a lasting impact on society, both by deepening our understanding of the universe and through direct contributions to technology and innovation. Yet, the complex nature of the field can make it challenging for non-scientists to follow and appreciate its advancements. Consequently, public engagement and the accessibility of information for the broader public have become significant priorities of the particle physics community. The ATLAS Collaboration embraces this responsibility, aiming to connect cutting-edge particle physics research with society.

Although in-person visits to the ATLAS detector and its control room are highly inspiring, access is often restricted by geographical distance, financial costs, physical ability, time limitations, and visa requirements. Consequently, many interested individuals worldwide are unable to participate in on-site visits. To overcome these barriers, the ATLAS Collaboration established the Virtual Visits programme~\cite{LeBoulicaut:2024vtl,Alhroob:2020ctk,Alhroob:2024xai,Manco:2022ryl,Mobius:2025fwh} as an experiment-level outreach initiative. It employs multimedia technology to bring particle physics research to global audiences, without the need to travel. The programme uses a combination of videoconferencing, video streaming, interactive question-and-answer sessions, and social media to boost its science outreach. While many video recordings, including a $360^{\circ}$ tour~\cite{360_tour} and a virtual reality tour~\cite{Lohwasser:2025dW}, are available online to inform remote audiences about ATLAS, CERN, and particle physics, the Virtual Visits were established to enable direct, real-time interaction between participants and ATLAS hosts. The hosts are trained members of the ATLAS Collaboration or CERN personnel, at any career stage, responsible for leading the Virtual Visits and answering audience questions. This innovative programme has evolved into an effective platform, allowing audiences worldwide to explore particle physics without geographical constraints.
Hundreds of virtual tours have connected thousands of participants from schools, community organisations, and public institutions worldwide, including those in remote or challenging locations such as the South Pole, refugee camps in Turkey, and detention centres in Greece. These tours have inspired school students and teachers, informed policymakers, and engaged the general public. Throughout the years, the Virtual Visits programme has adapted and expanded. Nevertheless, its objectives remain consistent: to reduce barriers, provide access to physics for global communities in their local languages, and contribute to worldwide science education and outreach efforts.

This paper describes and evaluates the key developments and impact of the programme. It begins with a review of the programme’s origins and evolution, followed by an analysis of its current global reach, supported by quantitative data. The statistics presented in this paper cover the period from January 2019 to December 2025, as systematic metadata collection only started at that time. The role of social media in expanding the visibility and engagement of Virtual Visits is then examined, with particular attention to recent trends. The impact of the programme on audiences is evaluated through participant feedback and illustrative case studies from Brazil and Greece. The paper concludes with a discussion of host training and engagement, providing a comprehensive overview of the structure and implementation of the programme.


%
\section{Concept, evolution and expansion}
%


The ATLAS Virtual Visits programme, launched in October of 2010, began as a practical solution to connect those unable to visit the ATLAS detector with researchers in particle physics, driven by a mission to reduce geographical barriers and stimulate interest worldwide. This initiative used emerging videoconferencing tools to provide an interactive experience. What started as an experimental outreach effort has since evolved into a significant component of the collaboration’s commitment to enhancing access to science.
\begin{figure}[!ht]
\centering
\includegraphics[width=.8\textwidth]{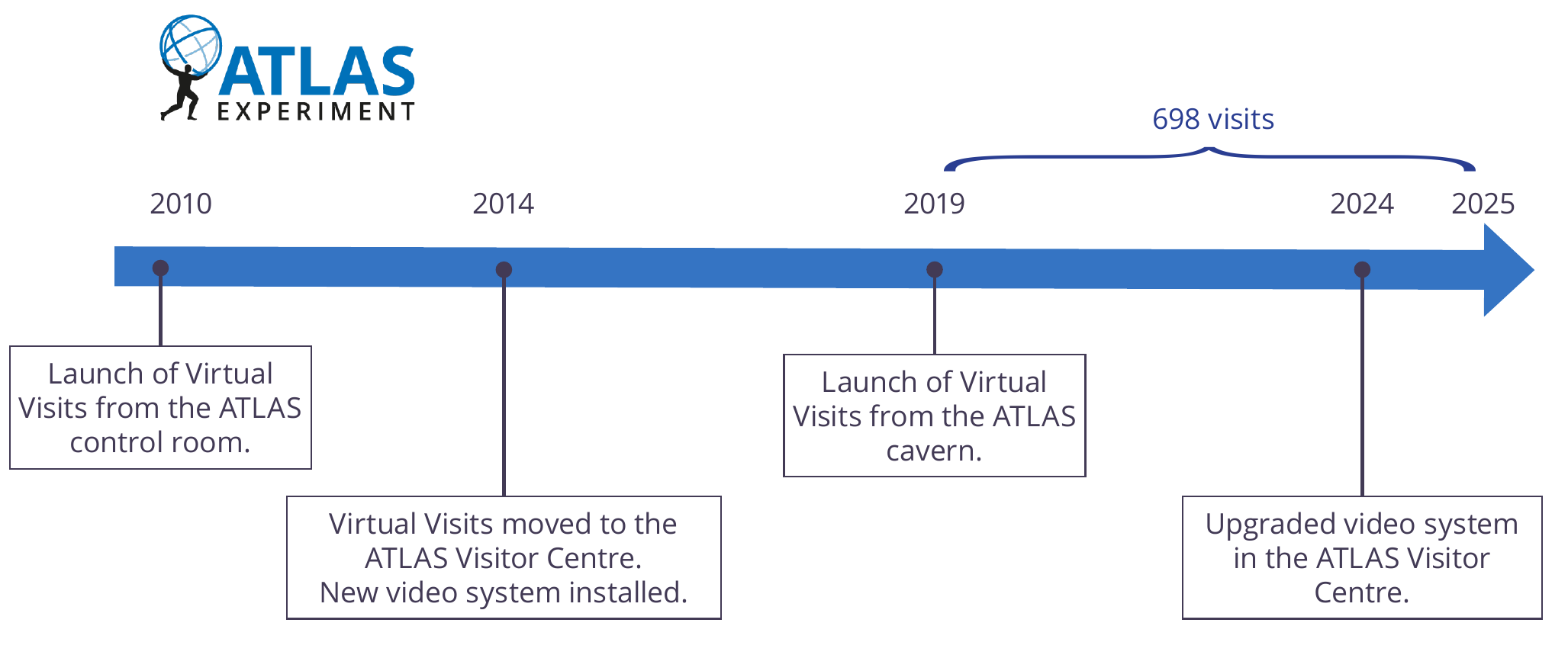}
\caption{Timeline showing key milestones in the evolution of the ATLAS Virtual Visits programme from 2010 to 2025. The figure highlights the launch of visits from the ATLAS control room, the relocation to the ATLAS Visitor Centre, the introduction of underground cavern visits in 2019, and subsequent system upgrades.}
\label{fig:timeline}
\end{figure}

Initially, ATLAS Virtual Visits were conducted from within the ATLAS control room, moving among the operational team~\cite{Goldfarb:2012fj}. As the frequency of visits increased, the broadcasting site was relocated to the ATLAS Visitor Centre (AVC) outside the control room to minimise disruptions to the detector’s daily operations. From this position, the visibility of the operational team is maintained through a large glass window separating the two areas. Two pre-installed high-definition cameras are used: one outside the control room focused on the hosts, and another inside providing a panoramic view of the control room and operational team. The videoconferencing system and associated video applications were upgraded multiple times to improve functionality and performance. A significant development occurred in 2019 with the introduction of underground visits, which take place in the ATLAS detector cavern~\cite{SanchezPineda:2019uiu}. These are only possible when the accelerator is not running. During maintenance and upgrade periods, hosts descend into the ATLAS experimental cavern, using smartphones to stream live views of the detector. The key milestones of the programme’s development are summarised in~\Cref{fig:timeline}.

Virtual Visits are booked via an online form~\cite{ATLAS_vv}, in which organisers indicate their desired date and time, the size and level of their group, the language, and, optionally, any other useful information. The visit is usually confirmed within a few days, depending on the availability of hosts.
A typical Virtual Visit lasts about an hour and begins with a welcome from the hosts, followed by an introduction to the key concepts relevant to the visit. These include the basics of particle physics, the structure and objectives of CERN, the LHC, and the ATLAS Experiment.
The introduction is followed by an explanation of the principles of particle detection and the structure of the ATLAS detector. When in the AVC, this is done with the help of detector parts on display in the room, and an ATLAS detector LEGO model~\cite{lego_models}. When underground, the host can directly point to the full detector systems, which are usually exposed due to the open configuration of the detector during shutdowns. After covering these technical aspects, the hosts discuss some of the major open questions in physics that the ATLAS Experiment is trying to answer. They may also discuss how technologies developed at CERN can benefit the public, such as medical imaging or the World Wide Web. Another topic of interest is the day-to-day functioning of the ATLAS control room. To further enrich the experience, hosts can use laptops to share supplementary media. This prepares the audience for an interactive question-and-answer session, encouraging active participation and engagement. Throughout the visit, the host may share personal stories or experiences, such as their career path, which helps the audience connect and relate to them. When appropriate, the host may also encourage participants to explore additional outreach and education opportunities beyond the scope of the Virtual Visits that involve ATLAS data or expertise. Such activities include Masterclasses~\cite{Borgelt:2024wbl}, organised by the International Particle Physics Outreach Group (IPPOG)~\cite{IPPOG_paper}, or ATLAS Open Data~\cite{OpenData}. Hosts and audiences can mutually agree to record visits for future reference or educational use.

Underground visits offer several benefits compared with visits from the AVC. First, hosts have the opportunity to describe the process of descending into the cavern by going through it on camera. In addition, the audience can view the detector from various perspectives, as the host has the freedom to move around the entire cavern. On the other hand, the AVC provides several advantages. The cavern is only available during LHC shutdowns, and accessible from 7:30 AM to 7:30 PM, excluding weekends, while the AVC remains available around the clock, enabling Virtual Visits to be scheduled essentially any time. In addition, during data taking, visitors can enjoy a live view of the working shift crew.

\begin{figure}[!ht]
\centering
\includegraphics[width=.6\textwidth]{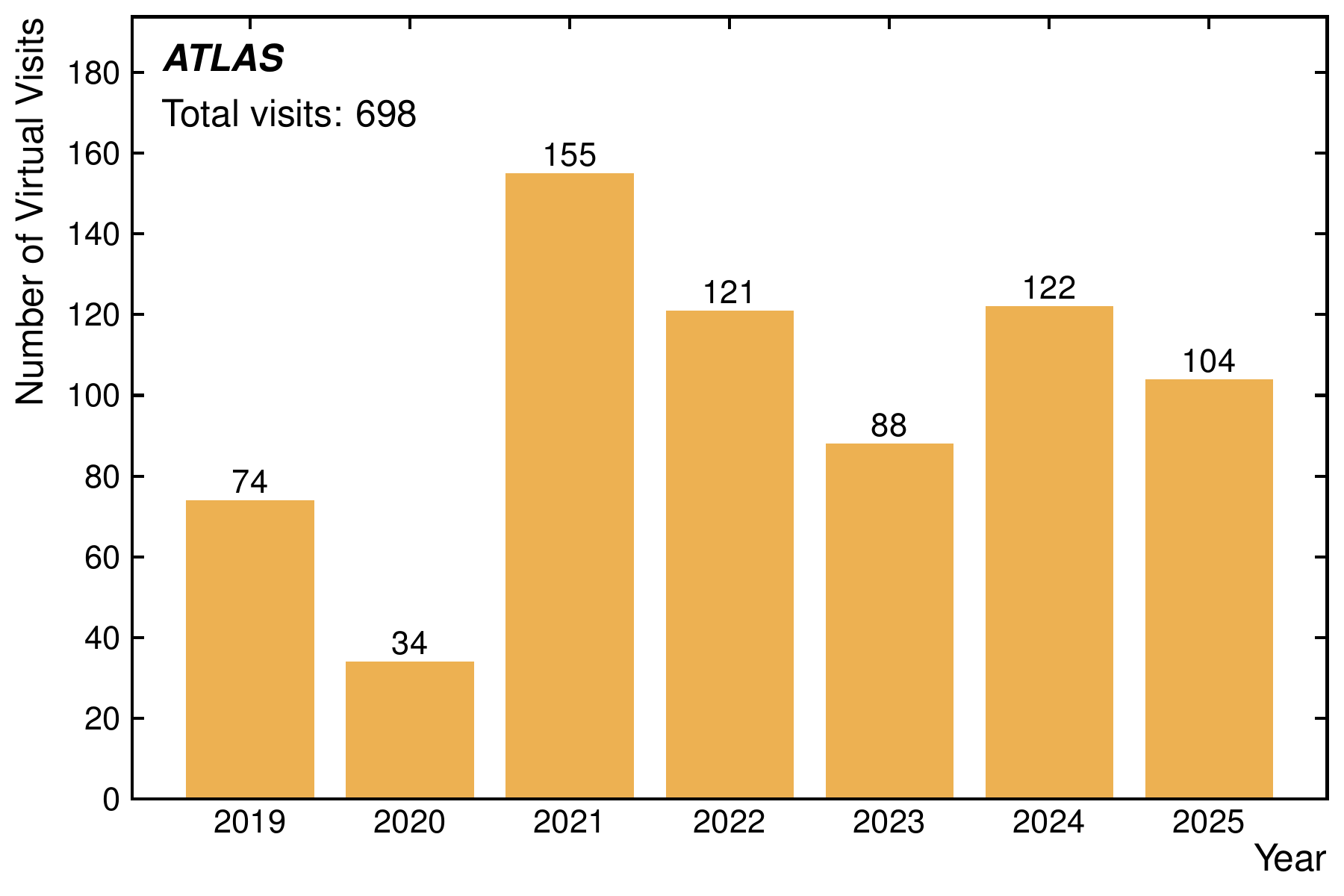}
\caption{Total number of ATLAS Virtual Visits per year from 2019 to 2025.}
\label{fig:total_visits}
\end{figure}

From a handful of visits in the first year, it now delivers over one hundred annually. With participant numbers per visit typically ranging from 10 to 50, and some exceptional visits reaching over 1,000 participants, the programme has engaged tens of thousands of individuals worldwide.~\Cref{fig:total_visits} illustrates the annual number of visits conducted between 2019 and 2025, totalling 698 visits, a testament to its sustained popularity. Two key trends are evident: a substantial decline in 2020, corresponding to the COVID-19 pandemic, and a marked rise in demand in 2021, attributed to the adoption of online videoconferencing and the temporary suspension of in-person visits at CERN.

\section{Participation and global reach}
%


The ATLAS Virtual Visits programme has expanded substantially since its inception, both in the number of visits and in its geographical reach. This section examines the programme’s scope, focusing on participation trends, regional engagement patterns, and a detailed country-level analysis supported by quantitative data and visual representations.

The primary participants are schools, universities, and community organisations, making up the majority of visits. For example, Virtual Visits are often used as an introduction or complement to Masterclasses, which helps strengthen both programmes. However, its reach also extends to more isolated and unconventional groups, such as refugee camps, Antarctic research stations, and detention centres, highlighting its role in overcoming physical and social barriers to scientific engagement. Virtual Visits have also been featured in science fairs and "researchers' nights" aimed at the general public. For example, Virtual Visits were featured in Researchers’ Night events organised by Italian groups in 2022, and in a science fair hosted by a German group on 4 July 2022, marking the $10^{\text{th}}$ anniversary of the Higgs boson discovery~\cite{higgs_discovery}. They have also accompanied public events during international conferences, such as the International Conference for High Energy Physics (ICHEP) in 2020, and during the ATLAS Collaboration Week in October 2025. In the United Kingdom, several Virtual Visits have been included in institutional open days, for instance, at the Rutherford Appleton Laboratory. For large visits of this type, an additional moderator, not onsite with the guides, is often needed to help chair the question-and-answer session.

In each case, the content is tailored to the specific audience. For school groups, visits emphasise accessible explanations of basic physics concepts and narratives about scientific careers. Sessions with university students address topics in greater depth, while discussions with engineers focus on detector design and operation. For computer scientists, the emphasis is on data acquisition, processing, and analysis. For policymakers and funding agencies, the focus is on project management, resource coordination, and the societal impact of fundamental research.

\subsection{Worldwide participation}
Since 2019, the ATLAS Virtual Visits programme has attracted participation from almost 70 countries, beyond ATLAS member states. The programme organises sessions accommodating participants from multiple countries who share linguistic ties, such as Portuguese-speaking individuals from Portugal or Angola joining visits organised by groups in Brazil, or Spanish-speaking participants from various Latin American countries joining sessions organised by groups in Spain and Mexico. This cross-border participation extends the programme's coverage and supports scientific interaction within linguistic communities.

\begin{figure}[!ht]
\begin{subfigure}[b]{0.48\textwidth}
\centering
\includegraphics[width=\textwidth]{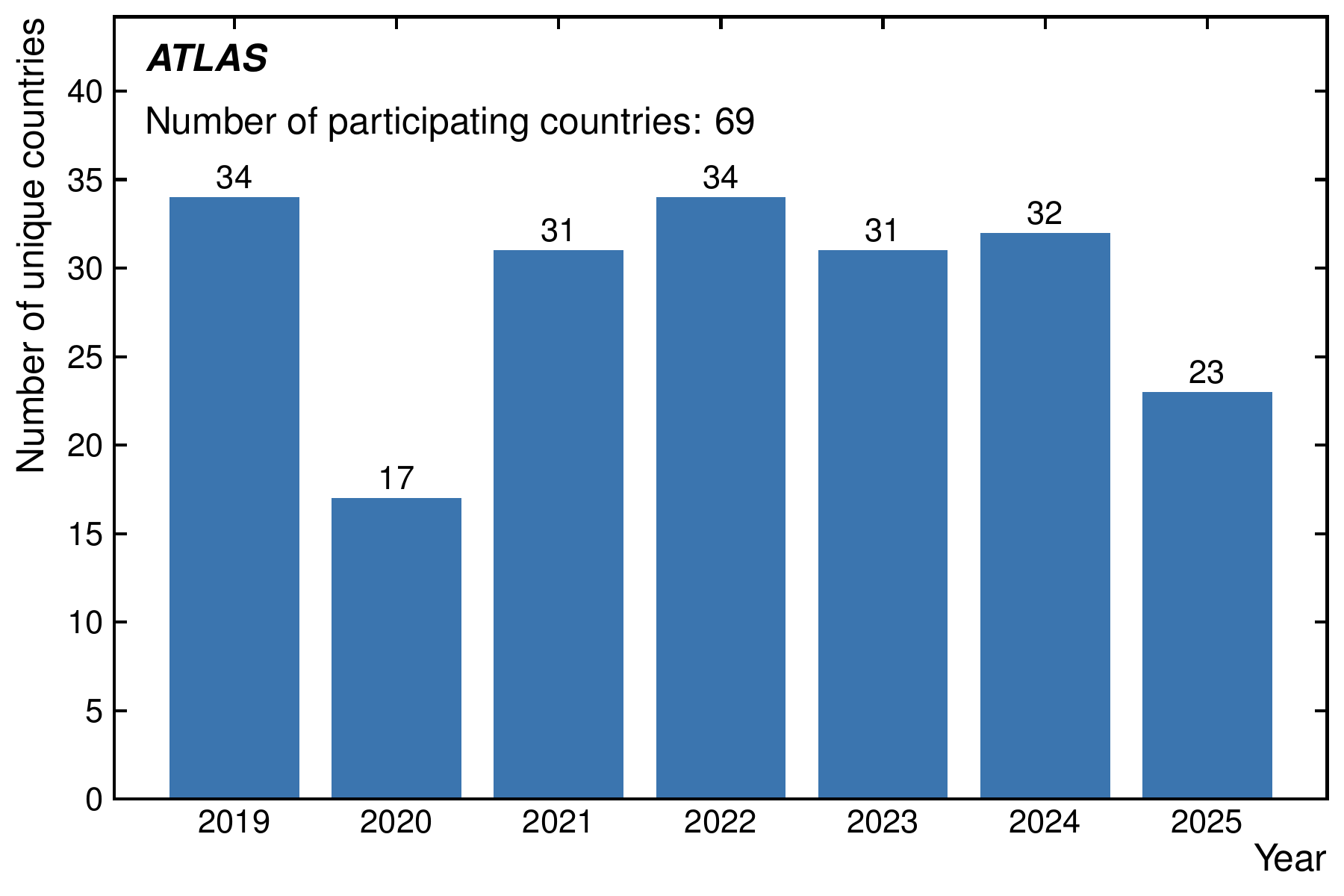}
\caption{}
\label{fig:countries_per_year}
\end{subfigure}
\begin{subfigure}[b]{0.48\textwidth}
\centering
\includegraphics[width=\textwidth]{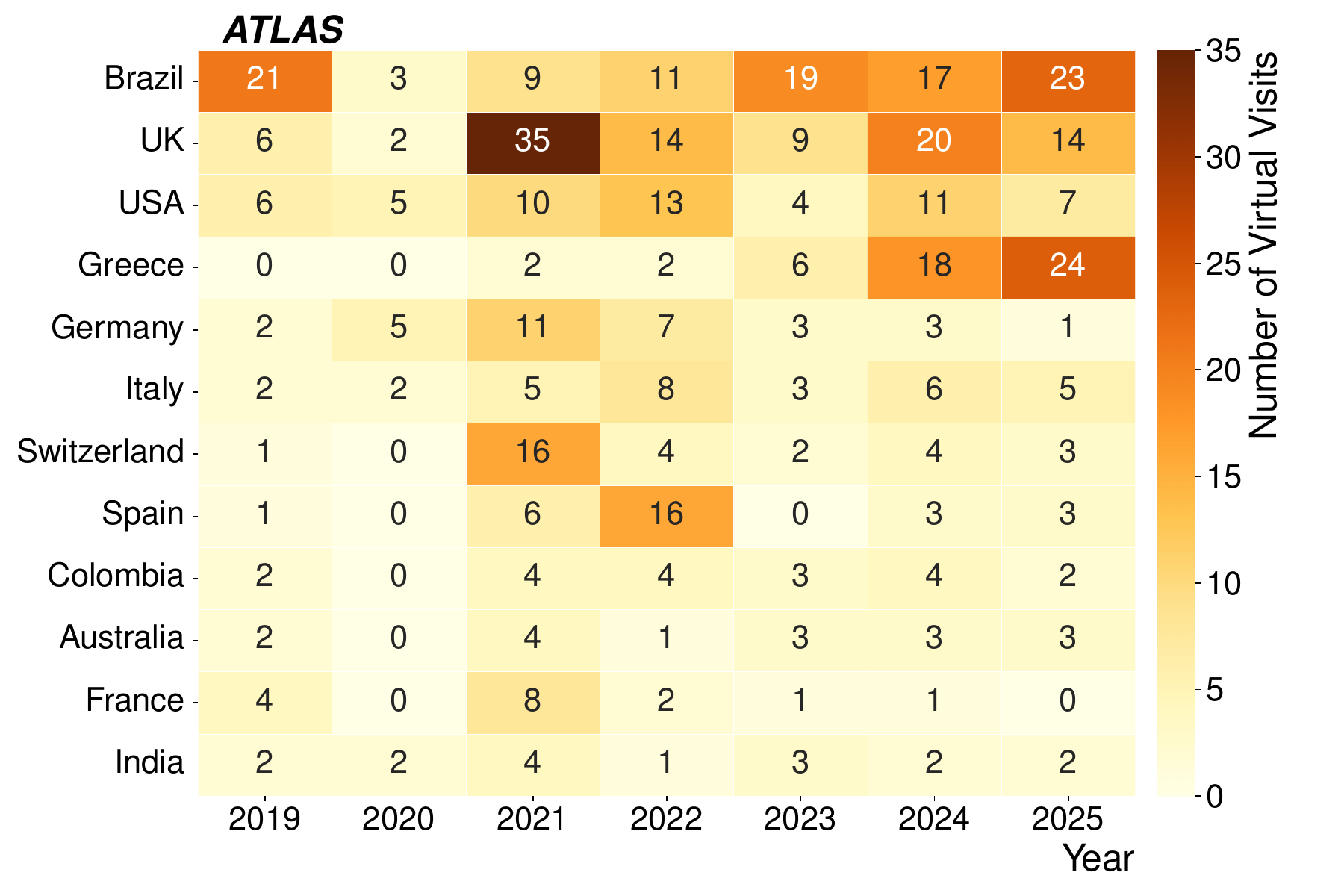}
\caption{}
\label{fig:country_heatmap}
\end{subfigure}

\caption{Country participation between 2019 and 2025: (a) number of countries per year and (b) number of visits per country per year for top participating countries.}
\end{figure}

The analysis of country participation reveals a broad reach and adaptability to diverse cultural contexts.~\Cref{fig:countries_per_year} presents the number of participating countries per year since 2019 and~\Cref{fig:country_heatmap} shows the number of visits per year for the most active countries. In cases where the visit included audiences from multiple countries speaking the same language, each country is counted for that visit. The top three countries, Brazil, the United Kingdom, and the United States of America, have all demonstrated sustained involvement, regularly incorporating Virtual Visits into high school and university curricula.

The fourth most active country, Greece, has seen a recent surge in participation, as described in~\Cref{sec:study_case}. Germany, Italy, Switzerland, Spain, and France all owe their higher participation to their familiarity with CERN. Australia, Colombia, and India have also recorded a significant number of visits, despite their physical distance.

A notable trend seen in~\Cref{fig:country_heatmap} is that many countries experienced their highest levels of participation in 2021, coinciding with the COVID-19 pandemic, when in-person visits were temporarily suspended, and virtual activities became more prominent. Some countries, including the United Kingdom, Brazil, the United States, and Italy, have exhibited relatively consistent levels of participation, excluding disruptions due to the pandemic. Greece, on the other hand, has shown a substantial increase in participation since 2021. For other countries, such as Spain and France, the number of visits peaks over a period of 1-2 years, before falling again. Many of these trends are correlated with the level of involvement of the hosts from a given country, the arrival and departure of particularly active hosts, and the degree to which they promote the programme in their home countries. This illustrates the importance of host engagement and strong networks.

\begin{figure}[!ht]
\centering
\includegraphics[width=.90\textwidth]{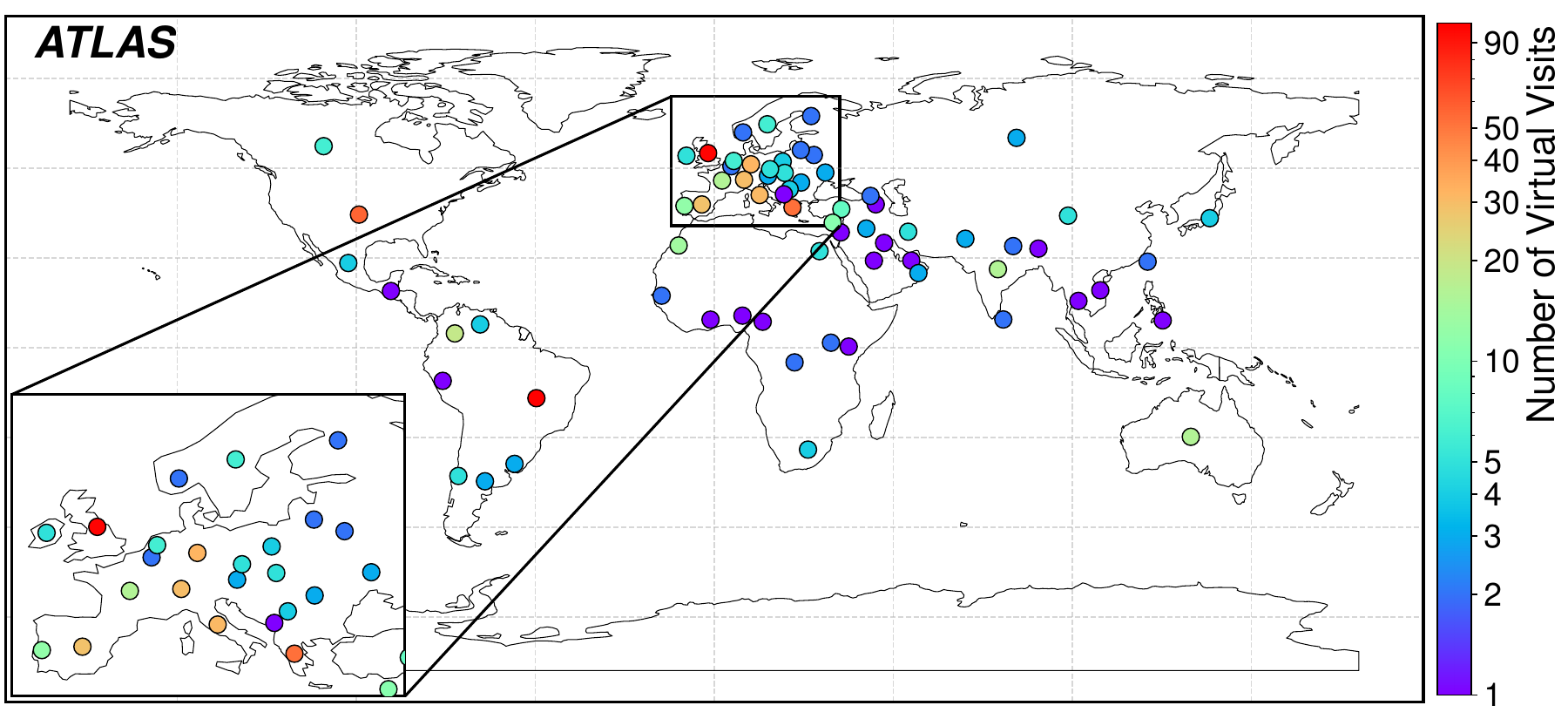}
\caption{World map showing the number of Virtual Visits for all participating countries between 2019 and 2025.}
\label{fig:countries_map}
\end{figure}

\Cref{fig:countries_map} shows a world map with each pin representing a country, coloured according to its total number of visits since 2019. This visualisation indicates that the programme’s reach extends well beyond the countries listed in~\Cref{fig:country_heatmap}. The regions with the largest fraction of participating countries in Virtual Visits are Western Europe and North America. This is likely due to the large number of ATLAS members working in these countries, which results in a greater awareness of the programme. South America also shows good coverage, with Brazil leading the way. Other countries, such as Colombia and Argentina, are also participating, partially due to the Masterclass programme. The Middle East has seen increased activity in recent years, particularly in Saudi Arabia and the United Arab Emirates.
Engagement in Asia is mainly driven by Japan and India, primarily through academic involvement. Despite some gaps in coverage, participation is emerging in Africa, with sessions, for example, in Ghana for school-aged students and in South Africa at the university level. Finally, Australia extends the reach of the programme to Oceania. Overall, these patterns of participation roughly reflect the distribution of ATLAS institutes. However, the programme still reaches countries with no universities or research institutes formally affiliated with the ATLAS Collaboration, such as India, Sri Lanka, Egypt and Iran. This is achieved by promoting the programme through personal and professional connections, and posters and presentations at international conferences.

\subsection{Languages and accessibility}
ATLAS Virtual Visits are offered in multiple languages to facilitate participation. Between 2019 and 2025, the programme employed 19 languages, enabling access to particle physics content for non-English-speaking audiences. The selection of languages reflects regional demand and the availability of hosts, supporting engagement in participants’ native languages.

\begin{figure}[!ht]
\begin{subfigure}[b]{0.48\textwidth}
\centering
\includegraphics[width=\textwidth]{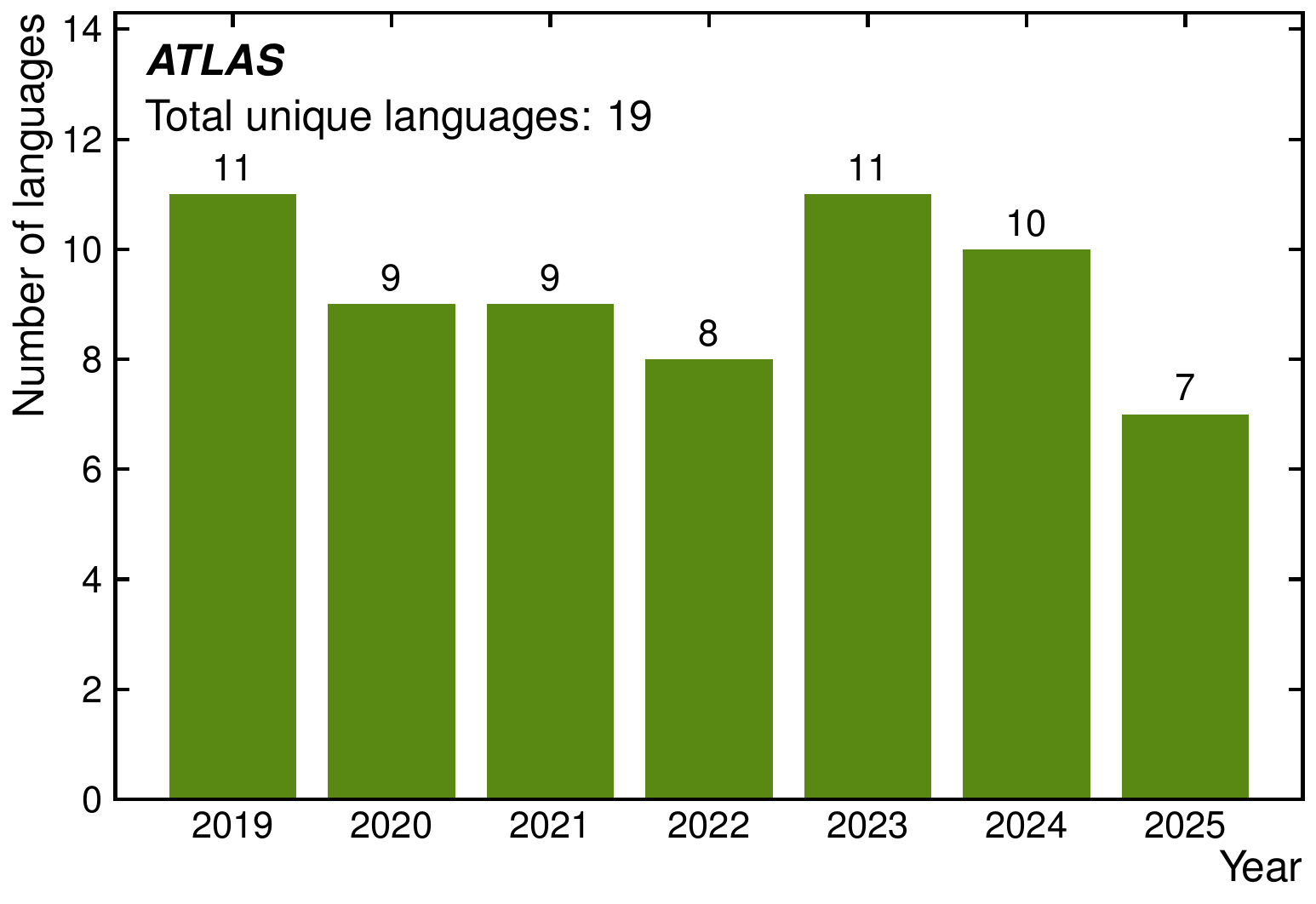}
\caption{}
\label{fig:language_per_year}
\end{subfigure}
\begin{subfigure}[b]{0.48\textwidth}
\centering
\includegraphics[width=\textwidth]{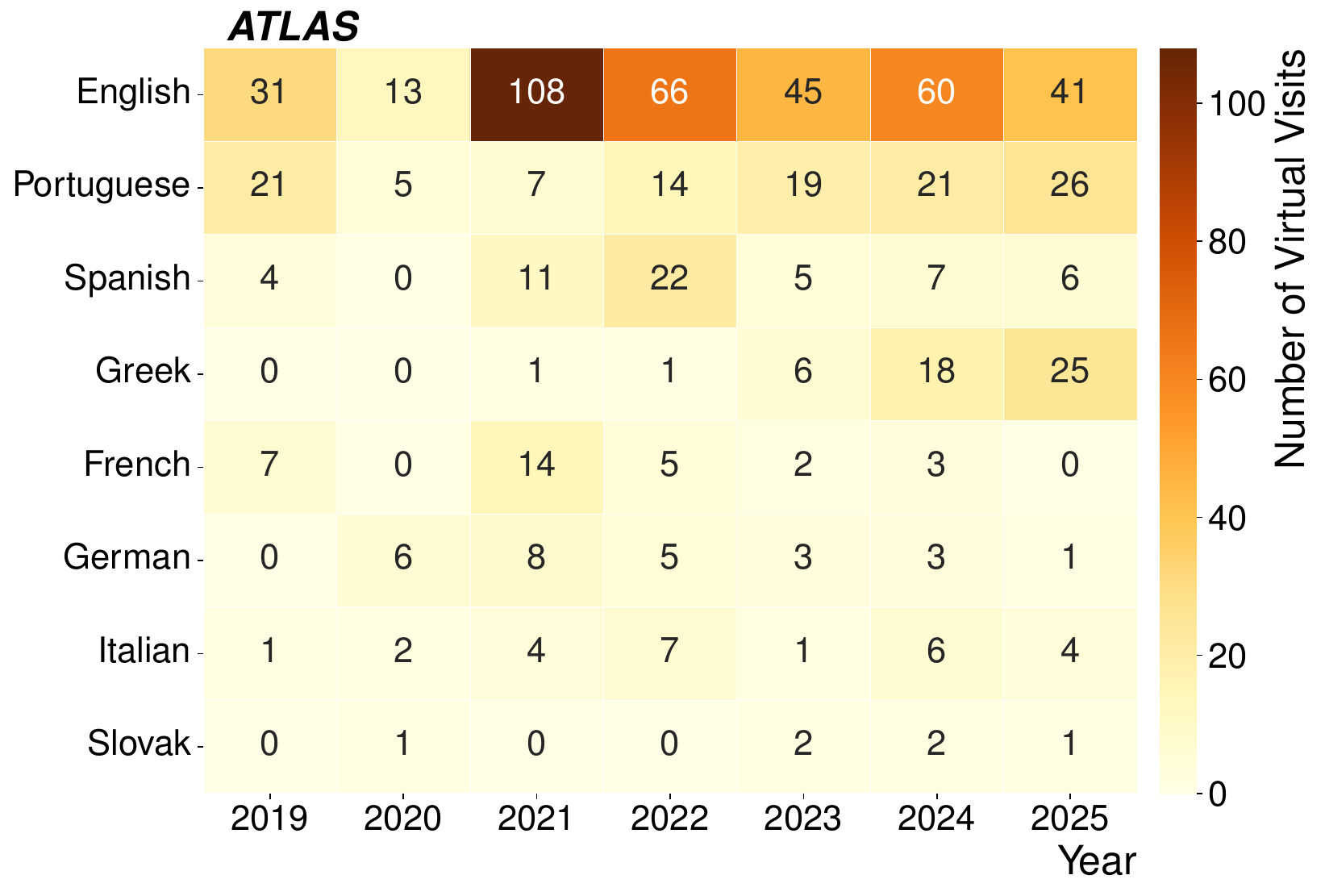}
\caption{}
\label{fig:language_heatmap}
\end{subfigure}

\caption{Language participation between 2019 and 2025: (a) number of languages per year and (b) number of visits per language per year for the most used languages.}
\end{figure}

\Cref{fig:language_per_year} shows the number of languages used in Virtual Visits per year, demonstrating a consistent diversity. ~\Cref{fig:language_heatmap} illustrates the temporal variations for the most commonly used languages. It shows that English was the most frequent, with 364 visits recorded in the last seven years, corresponding to 52\% of all visits. Portuguese accounted for 113 visits (16\%), primarily due to participation from Brazil, while Spanish, with 56 visits (8\%), facilitated engagement in Latin America. Other languages included are Greek, French, German, Italian, and Slovak.

The variations in time roughly follow the per-country variations shown in~\Cref{fig:country_heatmap}. English recorded a maximum of 108 visits in 2021, marking its highest annual usage. Portuguese, on the other hand, showed consistent levels of engagement throughout the observed period. The recent increase in Greek usage can also be noted, coming from audiences in Greece, Cyprus, and Greek schools abroad, e.g. in Germany or the Democratic Republic of the Congo (DRC). These patterns highlight the programme’s responsiveness to regional linguistic preferences.


\section{Open Virtual Visits and social media}
%


Typical Virtual Visits require a structured group of at least ten people, with one main organiser who books the visit and receives the video-conferencing details. The COVID-19 pandemic led to a significant increase in demand for Virtual Visits, particularly among individuals and small groups, with a peak observed in 2021. However, constraints related to staffing, scheduling, and time zones limited the capacity to accommodate such requests. In response, the ATLAS Collaboration launched Open Virtual Visits in the summer of 2021, featuring dedicated sessions specifically designed for individuals who are not part of an organised group. These sessions initially required pre-registration and were conducted on Zoom~\cite{Zoom}. Between 2021 and 2023, 15 visits took place on Zoom, with seven visits in 2021, seven in 2022, and one in 2023, at which point other platforms took precedence.

The ATLAS Collaboration maintains a strong presence on major social media platforms, including Instagram~\cite{instagram}, YouTube~\cite{youtube}, TikTok~\cite{tiktok} and Facebook~\cite{facebook}. These platforms are actively used to promote Open Virtual Visits. This includes static image posts, short-form videos, stories, and scheduled event reminders, where available. These social media campaigns are central in raising awareness and driving attendance for Virtual Visits, and ensuring broader engagement and reaching diverse audiences. To increase accessibility, Open Virtual Visits transitioned to live streaming, removing the need for pre-registration and making participation less intimidating. These visits are primarily streamed on YouTube and TikTok, enabling real-time interaction through chat features. Additionally, YouTube’s Video On Demand feature allows viewers to watch the live session later. While late viewers cannot participate in live discussions, they still benefit from hearing the questions asked by live viewers during the stream. Beyond the streams on the ATLAS YouTube channel, the programme's reach is extended by groups who record Virtual Visits and stream them on their platforms. For example, Virtual Visit hosts or organisers in Greece and Brazil have broadcast their sessions on their YouTube channels, in certain cases reaching thousands of viewers, significantly increasing the impact and accessibility of the events within their local communities.

\begin{figure}[!ht]
\centering
\includegraphics[width=.80\textwidth]{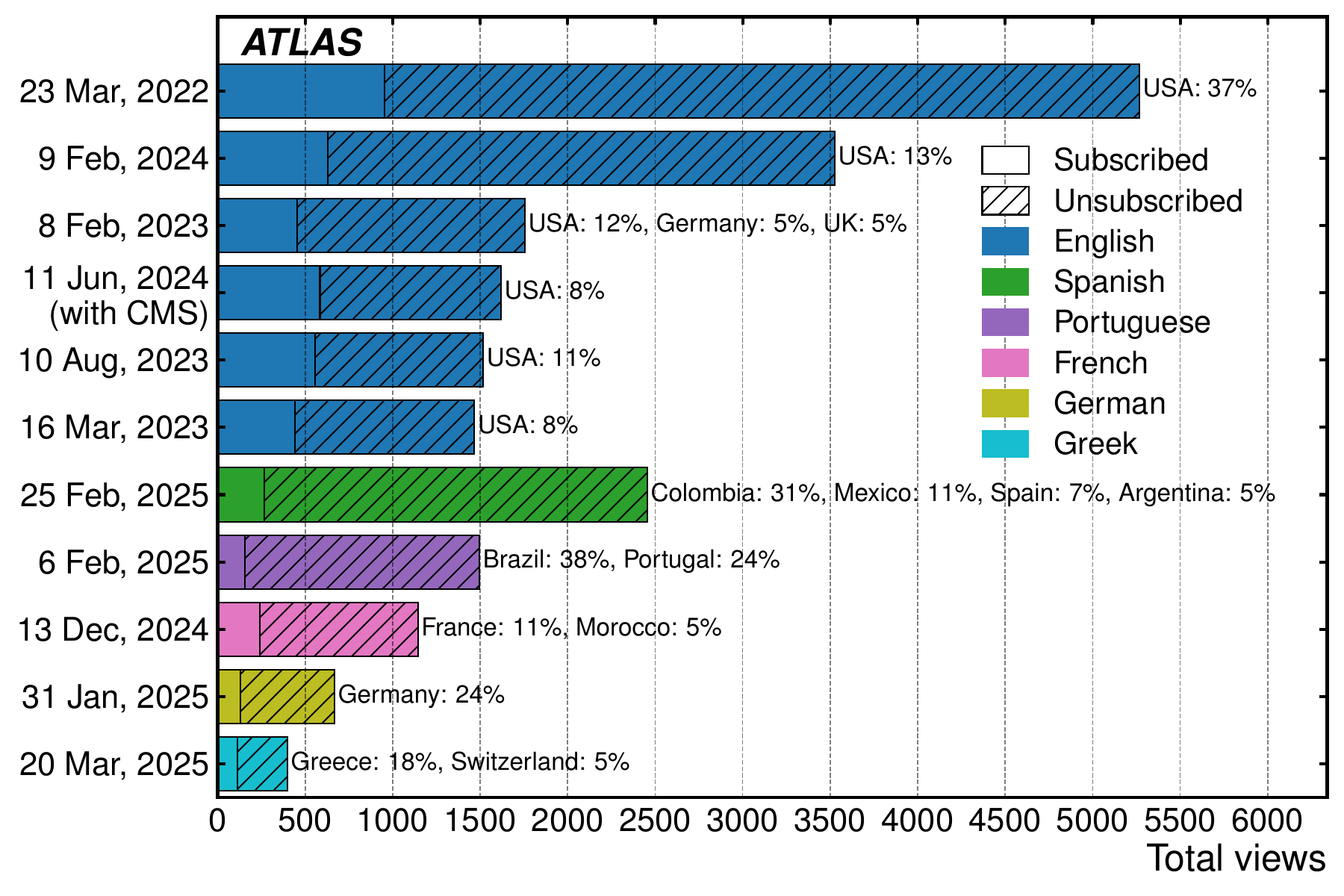}
\caption{Number of views, including both live and on-demand, for Open Virtual Visits streamed on the ATLAS YouTube channel, ordered by language and then by views within each language, as of December 2025. The text on the $y$-axis indicates the date each Virtual Visit was streamed, and the text to the right of each bar shows the percentage of viewers in each country. Given that this information is not available for all viewers, these numbers should be treated as minimum percentages. Only countries with more than 5\% of views are included. Viewer counts include both subscribed and unsubscribed users of the ATLAS YouTube channel. Data were obtained using Google's YouTube Analytics~\cite{youtube_analytics}.}
\label{fig:OVV}
\end{figure}

With a healthy backlog of English-language Open Virtual Visits on its YouTube channel in late 2024, ATLAS launched a new series of YouTube-based Virtual Visits in multiple languages, including French, Spanish, Greek, German, and Portuguese. This initiative aims to make the content more welcoming to non-English-speaking audiences and, guided by hosts who speak in their native languages, reflects the ATLAS Collaboration’s pluralism and cultural diversity~\cite{EDI_statement}. Since 2022, a total of 11 visits have been live streamed on YouTube, including five in languages other than English. The impact of this expansion is shown in~\Cref{fig:OVV}, which indicates that multilingual visits reach countries with limited viewership of the English-language videos and draw a higher fraction of viewers from countries where the visit language is spoken natively.


\section{Educational and outreach impact}
%

\subsection{The multi-faceted impact of Virtual Visits}

The success of the Virtual Visit programme is based on its ability to spark and foster engagement with the audience. Exploration around and inside the detector, or the views and descriptions of the control room, are clearly valuable from an educational point of view. More important are their effects on prompting questions and discussion from the audience, encouraging engagement with the hosts. Beyond that -- and arguably even more critical -- the personal stories offered by the hosts of their work and experiences in the collaboration help to build trust, understanding and empathy with the viewers. This is important for audiences considering careers in science and technology, and is key to encouraging those who might have pre-set prejudices to put a more human face on research and consider the value science brings to their lives.

\subsection{Examples of feedback on the impact}

From students and educators to community event organisers, participants’ feedback demonstrates the programme’s ability to inform, educate and connect diverse audiences with particle physics research. This section presents testimonials from visits conducted between January 2022 and December 2024, submitted through feedback forms or by email to hosts and the Virtual Visits team. Although this feedback is self-reported and anecdotal, it provides insight into the positive impact perceived by participants. Given that the primary goal of Virtual Visits is to inspire, these testimonies offer a particularly meaningful perspective on aspects of the programme’s impact that are not accessible through quantitative measures alone.

Most feedback comes from teachers, who often describe how Virtual Visits can be integrated into school activities and curricula. For example, a classroom in Sweden in March 2023 reported using a Virtual Visit as an introduction to a subsequent lecture on the Higgs boson. Similarly, a school in the USA in October 2024 launched a lecture series with a Virtual Visit, describing it as providing “wonderful insights into the science and technology utilised by ATLAS”. According to feedback from Greek teachers, Virtual Visits are “a unique, enriching, highly engaging and innovative teaching experience”. Several of them expressed an interest in integrating Virtual Visits into their curricula and advocated for their inclusion as a formal component of school programmes. A similar initiative has been realised in Australia's New South Wales state, where the Department of Education has, since 2017, built the resources to incorporate  Masterclasses coupled with Virtual Visits into high school curricula~\cite{Australia_curriculum}.

Teachers also frequently report that Virtual Visits motivate students' career aspirations in science. For example, following a visit in February 2022, a university professor in the USA shared that one of his students said, “In six more years, I could be doing that”, evidence that the visit had sparked a genuine sense of possibility. Similarly, in Northern Ireland, a teacher reported that a visit conducted in January 2023 encouraged interest in science, technology, engineering and mathematics (STEM) careers among his high school students, even though many were previously unaware of CERN. He noticed students exploring the CERN website after the visit -- a clear sign of their newfound interest.

Feedback received often highlights the sense of awe and excitement experienced by participants during the visit. For example, educators in Schools of Second Chance at Detention Centres in Greece described the experience as “a window into the world”. The Australian National Youth Science Forum books a Virtual Visit each January for their flagship programme, which engages hundreds of students from all over the country. In 2024, one student in the programme credited the session with inspiring their continued physics studies, while another was “enthralled” by antimatter measurements. A similar level of excitement was reached in a school in Germany in a September 2024 session, which was described as an “unprecedented success -- beyond anything [they] could have imagined”.

Another common thread was the continued engagement with CERN, ATLAS and particle physics prompted by the visit. In the USA, following a visit in October 2024, the organiser expressed enthusiasm for an in-person follow-up, noting how the virtual experience had inspired plans for a future visit to CERN. In India, in December 2024, the president of the non-profit organisation that initiated the Virtual Visit praised the session for “play[ing] a pivotal role in fostering young minds and motivating them to explore the realms of science and innovation”. Several groups have become regular participants, booking Virtual Visits every year, demonstrating their desire for long-term involvement.

Across regions, common observations emerge: the programme inspires engagement with science, stimulates curiosity and makes particle physics feel more accessible. Teachers emphasise its curricular relevance, while students express interest and motivation, with some considering careers at CERN. Organisers frequently acknowledge the hosts’ expertise, adaptability, and ability to address diverse audiences. Collectively, this feedback underscores ATLAS Virtual Visits as an effective educational tool, bridging cutting--edge research with audiences worldwide.


\section{Case studies of impactful ATLAS Virtual Visits}
\label{sec:study_case}


\subsection{Long-standing educational initiatives in Brazil}
Brazil’s participation in the ATLAS Virtual Visits programme highlights its role in making particle physics accessible across a diverse, geographically vast nation~\cite{seixas2019visitas}. Since 2012, Brazilian educators, inspired by CERN’s Portuguese Teachers’ Programme~\cite{teacher_programme}, have integrated Virtual Visits into school curricula and outreach events, engaging students from urban universities to rural communities. These sessions are often paired with International Masterclasses or university exhibitions, making them accessible to many students. A standout example is a session at the Federal University of Lavras in April 2025, which drew 380 students, exceeding the auditorium capacity.

Brazilian Virtual Visits place a strong emphasis on preparing students beforehand. This is typically done through a 40 -- 60 minute pre-visit introduction by an ATLAS member. In some cases, teachers encourage students to research CERN independently in the lead-up to the event. These strategies facilitate increased engagement, enabling students to participate in more comprehensive discussions of physics concepts, experimental methodologies, collaborative research models, career development, and the broader societal impacts of particle physics.

Despite challenges, such as inconsistent internet access limiting scalability across Brazil’s 30,000 high schools, the programme has made significant strides. Science clubs and public events expand access to underserved audiences, while prominent institutions, such as the University of São Paulo, integrate visits into STEM education. Feedback underscores the programme’s impact: a Lavras teacher noted that the 2025 session “sparked a passion for science”, with students expressing STEM career aspirations.

\subsection{New educational initiatives in Greece}
Greece’s surge in ATLAS Virtual Visits since 2023 demonstrates the programme’s capacity to deliver inclusive science education. Following requests from participants in the CERN Greek Teachers Programme, the initiative was launched in 2023 during the annual maintenance period of the LHC. From November 2023 to February 2024, 15 visits were conducted, engaging 430 schools and 13,000 participants. Continued demand led to 17 visits during the 2024 -- 2025 LHC shutdown, reaching 640 Greek-speaking schools across Greece, Cyprus, Germany, and the DRC, with close to 25,000 participants. In 2025, roughly 20\% of groups were returning from previous years.

The programme’s strength lies in its adaptability, serving kindergartens, secondary schools, special education institutions, and Schools of Second Chance in Detention Centres (SSCDCs). Although these groups were initially beyond the intended scope of the programme, their requests for Virtual Visits emphasised the importance of equal access, diversity, and inclusion, making a compelling case for participation. For example, secondary schools in remote regions, such as small islands, typically have limited access to scientific experiences and therefore value Virtual Visits as a unique opportunity. For primary school sessions, the content was adapted with age-appropriate tools such as colouring books. Most of the SSCDC participants demonstrated strong engagement and posed insightful questions. A session held in January 2025 with 50 schools achieved ``unprecedented'' engagement, with teachers noting sustained student focus -- a rarity for hour-long events.

The participants' response to the programme is overwhelmingly positive. Younger students submitted drawings following the visit, demonstrating enthusiasm and curiosity. For young children, ATLAS is a “scientists’ playground”; for older students, it is a technological marvel.


\section{Training and engagement of hosts}
%


The success of Virtual Visits depends on technology and on the individuals who bring it to life: the hosts. From the programme’s early days, it became clear that this initiative offered dual benefits: extending the reach of science communication while helping ATLAS Collaboration members further develop outreach and communication skills. With thousands of members and a wide range of expertise, the collaboration provides a large pool of hosts who connect deeply with global communities. Effectively matching hosts with audiences is essential, as requests from schools, community groups, and institutions often specify language preferences, educational levels, or scientific interests. Hosts are identified based on their language skills, availability, and, when possible, regional or institutional connections with the audience. In certain cases, relevant subject-matter expertise or gender, for example, for events celebrating women in science, are also considered.

The recruitment and training process for hosts has been adapted to meet these needs, balancing the selection of participants with relevant expertise and communication skills with opportunities for further skill development and accessible participation. The process begins by identifying suitable candidate hosts and providing training to support their growth. Hosting Virtual Visits is done on a volunteer basis. Unlike the formal training required for in-person ATLAS guides, the process for becoming a Virtual Visit host is simple and accessible. Volunteers are first briefed about navigating the ATLAS cavern and standard safety procedures. Then, they typically observe or co-host alongside experienced volunteers to build confidence through practical experience.

This approach rapidly equips new guides with valuable communication skills, making the Virtual Visit programme an effective introduction to outreach. Over time, many participants join a growing roster of guides and also volunteer for other ATLAS outreach initiatives, thereby extending the programme's impact within the collaboration. Perhaps more importantly, by working with their peers on visits, guides gain a broader perspective of how their research fits into the overall framework of society. As each guide has their own stories to tell and their own analogies to share, they learn from each other's viewpoints and become better, more well-rounded scientists in the process.

\begin{figure}[!ht]
\begin{subfigure}[b]{0.48\textwidth}
\centering
\includegraphics[width=\textwidth]{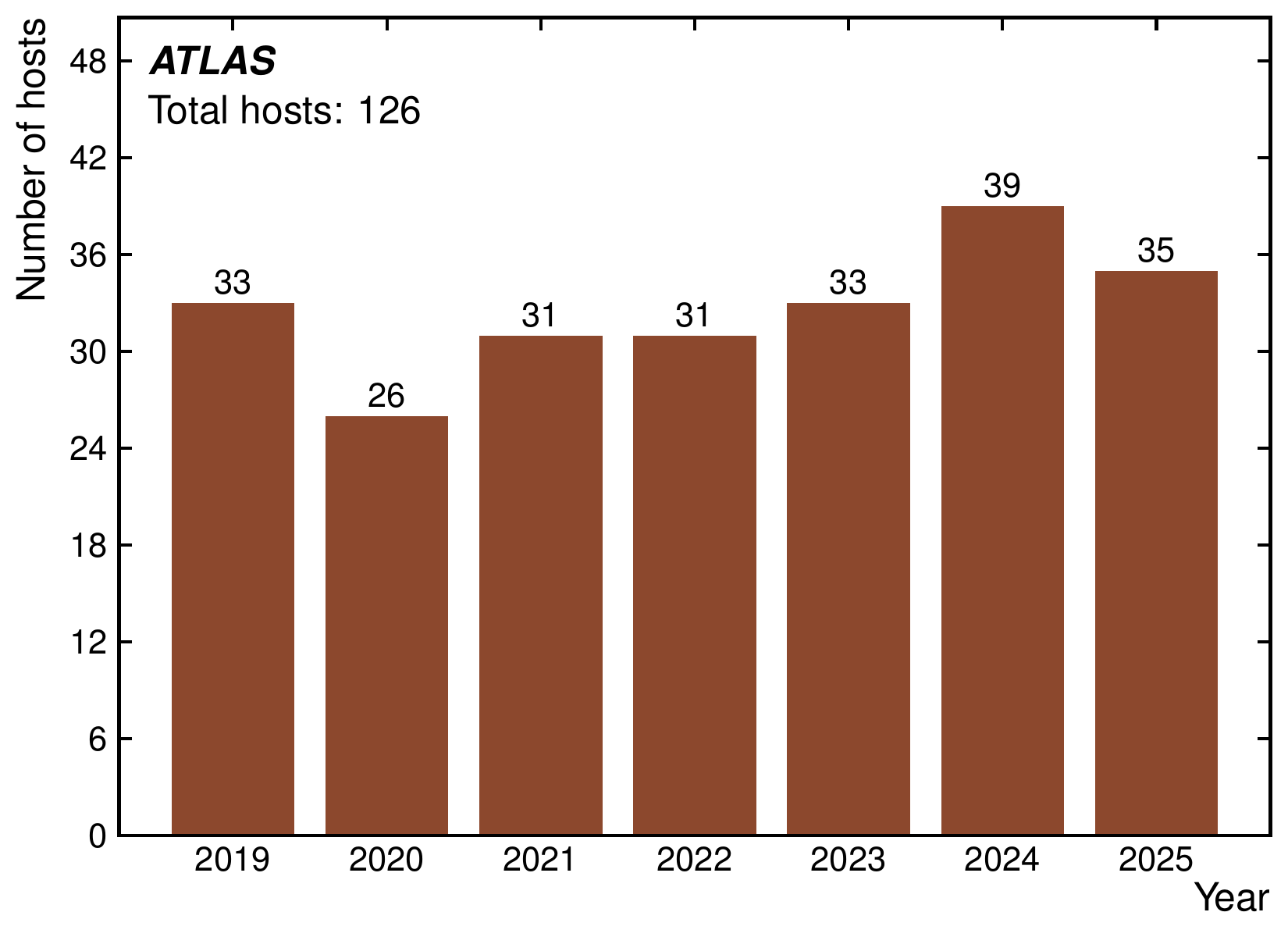}
\caption{}
\label{fig:hosts_per_year}
\end{subfigure}
\begin{subfigure}[b]{0.48\textwidth}
\centering
\includegraphics[width=\textwidth]{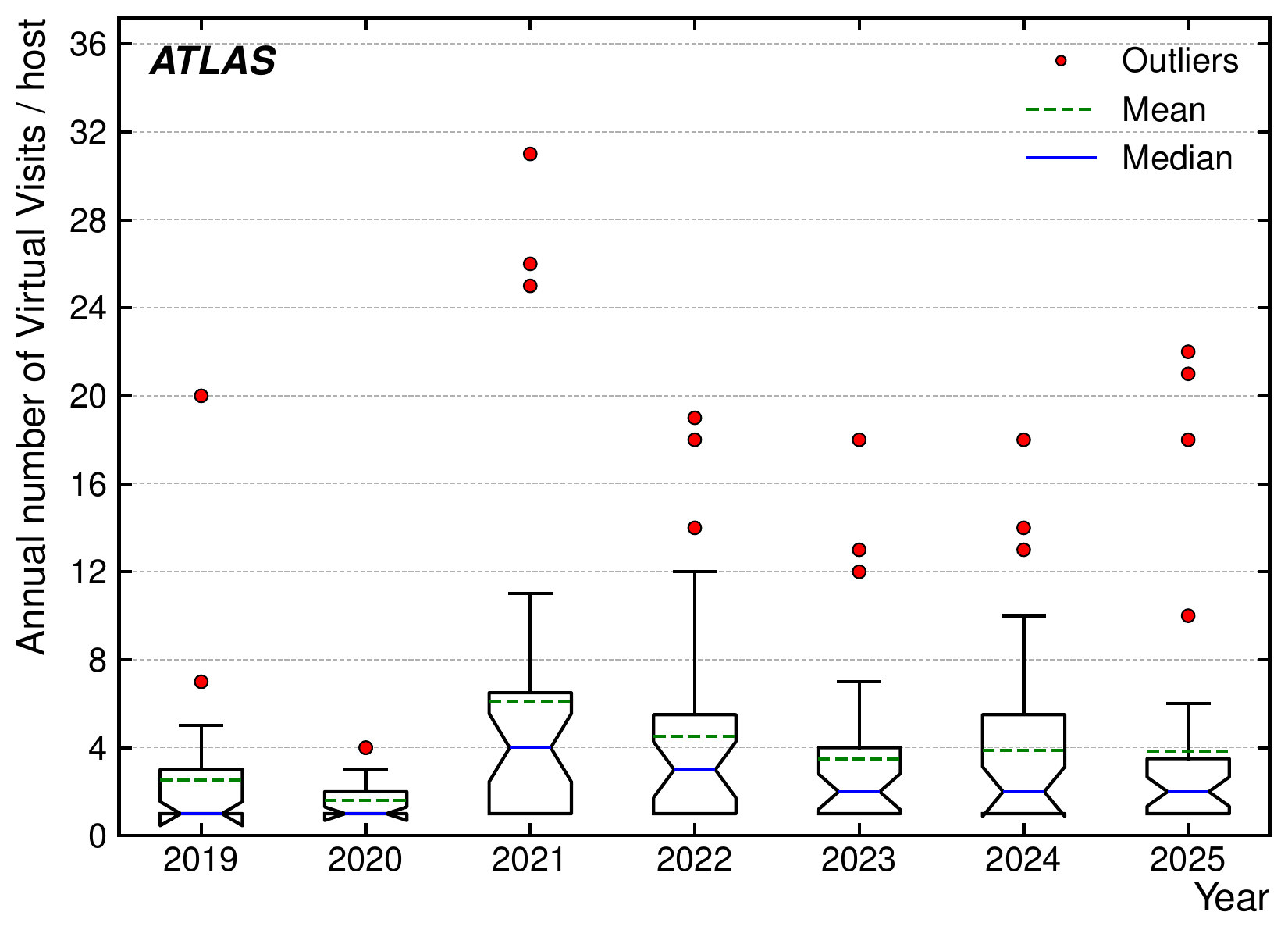}
\caption{}
\label{fig:host_average}
\end{subfigure}

\caption{Participating Virtual Visit hosts per year. Figure (a) shows the number of hosts participating annually, while Figure (b) shows the box diagram of the annual number of visits per host. The box spans the first ($25^{\text{th}}$ percentile) and third ($75^{\text{th}}$ percentile) quartiles. The central blue solid line marks the median, and the dashed green line shows the mean. Whiskers denote the minimum and maximum values within $1.5$ times the interquartile range, while solid red circles represent outliers beyond this range.}
\end{figure}

The data reflect sustained participation: over the past seven years, 126 individuals have actively hosted Virtual Visits, ensuring a reliable pool of volunteers for the expanding programme.~\Cref{fig:hosts_per_year} shows the number of participating hosts annually from 2019 to 2025, indicating stable participation throughout this period. However, host engagement levels vary, as shown in~\Cref{fig:host_average}.
This figure tracks the average number of visits conducted per host each year. The solid red circles represent highly active hosts who conduct significantly more visits than the median, suggesting that a core group is responsible for maintaining the programme's momentum.


\section{Future outlook}
%


After 15 years of operation, the ATLAS Virtual Visits programme will continue to build upon its previous success and address current limitations. For example, matching hosts to participant groups can be difficult when the requested language is less commonly spoken, which limits the pool of available hosts to only a few individuals. In addition, the international and academic nature of the ATLAS Collaboration implies that researchers are often based at CERN for a limited time, making host retention challenging. Despite these obstacles, the programme's core mission, connecting particle physics with curious minds, remains unchanged. To continually improve the programme, plans are underway to harness emerging technologies, expand engagement with new audiences, and diversify content.

One focus is on integrating new tools to enhance the virtual experience. For example, emerging translation tools driven by artificial intelligence could eliminate language barriers, enabling real-time multilingual sessions. This would help with Virtual Visit requests in less widely spoken languages, for which it is sometimes difficult to find an available host. Expanding audience reach is also a priority. While Virtual Visits have already engaged with unconventional groups, future efforts aim to expand access to underserved regions, such as rural communities in Africa and Asia, where the programme's reach remains limited. Collaborations with local organisations can facilitate this growth, tailoring content to regional needs and curricula. At the same time, increasing capacity is essential to meet the demand. Plans include sustaining the current annual benchmark of over 100 visits, supported by a larger and more diverse pool of trained hosts.

The Virtual Visits team also envisions leveraging additional ATLAS facilities beyond the AVC and underground cavern, such as laboratories where future ATLAS detector components are built and tested. These venues can offer fresh perspectives, showcasing different facets of the experiment and engaging audiences with different interests. They may also be able to accommodate different time zones more easily than the ATLAS underground cavern, which operates on a fixed schedule for safety reasons. Finally, this would allow ATLAS researchers based outside CERN to participate as hosts, increasing the size and diversity of the volunteer pool. Another way to vary the perspective is to carry out joint visits with other CERN experiments, such as CMS~\cite{CMS-CMS-00-001}, ALICE~\cite{ALICE:2008ngc}, and LHCb~\cite{LHCb:2008vvz}, or with other international experiments outside of CERN. A few of these have been conducted so far, for example, with CMS~\cite{ATLASVirtualVisitsTeam:2015kif}, and further opportunities are being explored. These collaborative sessions provide a global view of particle physics research at CERN, highlighting complementary efforts and promoting a deeper public understanding of the field.

This forward-looking vision hinges on continued adaptability, a hallmark of the programme’s success. By embracing innovations and expanding the programme, the aim is to inspire the next generation of scientists and enthusiasts, regardless of location or background.


\section{Conclusion}
%


The ATLAS Virtual Visits programme has become a central component of the collaboration's global science communication effort. Through live interaction, multilingual accessibility, and digital platforms, the programme has connected thousands of participants worldwide with ATLAS particle physics research. Since systematic data collection began in 2019, close to 700 visits have taken place with groups from nearly 70 countries, using 19 languages. Feedback from participants underscores the programme's educational impact, fostering curiosity and motivating young audiences to pursue careers in science and technology. The initiative also enables ATLAS members to enhance their communication skills and engage with diverse communities. Building on these achievements, the Virtual Visits team aims to expand the programme's reach further and adopt new technologies. These efforts play an important role in advancing the ATLAS Collaboration’s mission of connecting frontier particle physics research with society.


%
%
%
%
%
\section*{Acknowledgements}
%

%

%
%

%
%

We thank CERN for the very successful operation of the LHC and its injectors, as well as the support staff at
CERN and at our institutions worldwide without whom ATLAS could not be operated efficiently.

The crucial computing support from all WLCG partners is acknowledged gratefully, in particular from CERN, the ATLAS Tier-1 facilities at TRIUMF/SFU (Canada), NDGF (Denmark, Norway, Sweden), CC-IN2P3 (France), KIT/GridKA (Germany), INFN-CNAF (Italy), NL-T1 (Netherlands), PIC (Spain), RAL (UK) and BNL (USA), the Tier-2 facilities worldwide and large non-WLCG resource providers. Major contributors of computing resources are listed in Ref.~\cite{ATL-SOFT-PUB-2026-001}.

We gratefully acknowledge the support of ANPCyT, Argentina; YerPhI, Armenia; ARC, Australia; BMWFW and FWF, Austria; ANAS, Azerbaijan; CNPq and FAPESP, Brazil; NSERC, NRC and CFI, Canada; CERN; ANID, Chile; CAS, MOST and NSFC, China; Minciencias, Colombia; MEYS CR, Czech Republic; DNRF and DNSRC, Denmark; IN2P3-CNRS and CEA-DRF/IRFU, France; SRNSFG, Georgia; BMFTR, HGF and MPG, Germany; GSRI, Greece; RGC and Hong Kong SAR, China; ICHEP and Academy of Sciences and Humanities, Israel; INFN, Italy; MEXT and JSPS, Japan; CNRST, Morocco; NWO, Netherlands; RCN, Norway; MNiSW, Poland; FCT, Portugal; MNE/IFA, Romania; MSTDI, Serbia; MSSR, Slovakia; ARIS and MVZI, Slovenia; DSI/NRF, South Africa; MICIU/AEI, Spain; SRC and Wallenberg Foundation, Sweden; SERI, SNSF and Cantons of Bern and Geneva, Switzerland; NSTC, Taipei; TENMAK, T\"urkiye; STFC/UKRI, United Kingdom; DOE and NSF, United States of America.

Individual groups and members have received support from BCKDF, CANARIE, CRC and DRAC, Canada; CERN-CZ, FORTE and PRIMUS, Czech Republic; COST, ERC, ERDF, Horizon 2020 and Marie Sk{\l}odowska-Curie Actions, European Union; Investissements d'Avenir Labex, Investissements d'Avenir Idex and ANR, France; DFG and AvH Foundation, Germany; Herakleitos, Thales and Aristeia programmes co-financed by EU-ESF and the Greek NSRF, Greece; BSF-NSF and MINERVA, Israel; NCN and NAWA, Poland; La Caixa Banking Foundation, CERCA and AGAUR programs from Generalitat de Catalunya and PROMETEO and GenT Programmes Generalitat Valenciana, Spain; G\"{o}ran Gustafssons Stiftelse, Sweden; The Royal Society and Leverhulme Trust, United Kingdom; Eric and Wendy Schmidt Fund for Strategic Innovation, United States of America.

In addition, individual members wish to acknowledge support from Chile: Agencia Nacional de Investigaci\'on y Desarrollo (ANID FONDECYT reg. 1230987, FONDECYT 1230812, FONDECYT 1240864, Fondecyt 3240661, Fondecyt Regular 1240721); China: Chinese Ministry of Science and Technology (MOST-2023YFA1605700, MOST-2023YFA1609300), National Natural Science Foundation of China (NSFC - 12175119, NSFC 12275265); Czech Republic: Czech Science Foundation (GACR - 24-11373S), Ministry of Education Youth and Sports (ERC-CZ-LL2327, FORTE CZ.02.01.01/00/22\_008/0004632), PRIMUS Research Programme (PRIMUS/21/SCI/017); EU: H2020 European Research Council (ERC - 101002463); European Union: European Research Council (BARD No. 101116429, ERC - 948254, ERC 101089007), European Regional Development Fund (HE COFUND GA No.101081355, ERDF), European Union, Future Artificial Intelligence Research (FAIR-NextGenerationEU PE00000013), Marie Sklodowska-Curie Actions (GAP-101168829); France: Agence Nationale de la Recherche (ANR-21-CE31-0013, ANR-21-CE31-0022, ANR-22-EDIR-0002, ANR-24-CE31-0504-01); Germany: Deutsche Forschungsgemeinschaft (DFG - 469666862); China: Research Grants Council (GRF); Italy: Ministero dell'Università e della Ricerca (NextGenEU 153D23001490006 M4C2.1.1, NextGenEU I53D23000820006 M4C2.1.1, NextGenEU I53D23001490006 M4C2.1.1, SOE2024\_0000023); Japan: Japan Society for the Promotion of Science (JSPS KAKENHI  JP25H0063, JSPS KAKENHI JP22H01227, JSPS KAKENHI JP22H04944, JSPS KAKENHI JP22KK0227, JSPS KAKENHI JP24K23939, JSPS KAKENHI JP24KK0251, JSPS KAKENHI JP25H00650, JSPS KAKENHI JP25H01291, JSPS KAKENHI JP25K01023); Norway: Research Council of Norway (RCN-314472); Poland: Ministry of Science and Higher Education (IDUB AGH, POB8, D4 no 9722), Polish National Science Centre (NCN 2021/42/E/ST2/00350, NCN OPUS 2023/51/B/ST2/02507, NCN OPUS nr 2022/47/B/ST2/03059, NCN UMO-2019/34/E/ST2/00393, UMO-2022/47/O/ST2/00148, UMO-2023/49/B/ST2/04085, UMO-2023/51/B/ST2/00920, UMO-2024/53/N/ST2/00869); Spain: Agència de Gestió d'Ajuts Universitaris i de Recerca. (AGAUR - 2023 BP 00141), Generalitat Valenciana (ASFAE/2022/008), Ministry of Science and Innovation (RYC2019-028510-I, RYC2020-030254-I, RYC2021-031273-I, RYC2022-038164-I), Ministerio de Ciencia, Innovación y Universidades/Agencia Estatal de Investigaci\'on (PID2022-142604OB-C22); Sweden: Carl Trygger Foundation (Carl Trygger Foundation CTS 22:2312), Swedish Research Council (Swedish Research Council 2023-04654, VR 2021-03651, VR 2022-03845, VR 2022-04683, VR 2023-03403, VR 2024-05451), Knut and Alice Wallenberg Foundation (KAW 2018.0458, KAW 2023.0366); Switzerland: Swiss National Science Foundation (SNSF - PCEFP2\_194658); United Kingdom: The Binks Trust, Royal Society (NIF-R1-231091); United States of America: U.S. Department of Energy (ECA DE-AC02-76SF00515), John Templeton Foundation (John Templeton Foundation 63206), Neubauer Family Foundation.

%
%


%
\clearpage
\printbibliography
\clearpage
\input{atlas_authlist}
%

%
%

%

%
%
%
%

\end{document}

%% file: atlas_authlist.tex
 
\begin{flushleft}
\hypersetup{urlcolor=black}
{\Large The ATLAS Collaboration}

\bigskip

\AtlasOrcid[0000-0002-6665-4934]{G.~Aad}$^\textrm{\scriptsize 102}$,
\AtlasOrcid[0000-0001-7616-1554]{E.~Aakvaag}$^\textrm{\scriptsize 17}$,
\AtlasOrcid[0000-0002-5888-2734]{B.~Abbott}$^\textrm{\scriptsize 121}$,
\AtlasOrcid[0000-0002-0287-5869]{S.~Abdelhameed}$^\textrm{\scriptsize 83b}$,
\AtlasOrcid[0000-0002-1002-1652]{K.~Abeling}$^\textrm{\scriptsize 54}$,
\AtlasOrcid[0000-0001-5763-2760]{N.J.~Abicht}$^\textrm{\scriptsize 48}$,
\AtlasOrcid[0000-0002-8496-9294]{S.H.~Abidi}$^\textrm{\scriptsize 30}$,
\AtlasOrcid[0009-0003-6578-220X]{M.~Aboelela}$^\textrm{\scriptsize 44}$,
\AtlasOrcid[0000-0002-9987-2292]{A.~Aboulhorma}$^\textrm{\scriptsize 36e}$,
\AtlasOrcid[0000-0001-5329-6640]{H.~Abramowicz}$^\textrm{\scriptsize 154}$
% ...
\bigskip
\\

% ...
$^{17}$Department for Physics and Technology, University of Bergen, Bergen; Norway.\\
% ...
$^{30}$Physics Department, Brookhaven National Laboratory, Upton NY; United States of America.\\
% ...
$^{36}$$^{(a)}$Facult\'e des Sciences Ain Chock, Universit\'e Hassan II de Casablanca;$^{(b)}$Facult\'{e} des Sciences, Universit\'{e} Ibn-Tofail, K\'{e}nitra;$^{(c)}$Facult\'e des Sciences Semlalia, Universit\'e Cadi Ayyad, LPHEA-Marrakech;$^{(d)}$LPMR, Facult\'e des Sciences, Universit\'e Mohamed Premier, Oujda;$^{(e)}$Facult\'e des sciences, Universit\'e Mohammed V, Rabat;$^{(f)}$Institute of Applied Physics, Mohammed VI Polytechnic University, Ben Guerir; Morocco.\\
% ...
$^{44}$Physics Department, Southern Methodist University, Dallas TX; United States of America.\\
% ...
$^{48}$Fakult\"{a}t Physik, Technische Universit{\"a}t Dortmund, Dortmund; Germany.\\
% ...
$^{54}$II. Physikalisches Institut, Georg-August-Universit\"{a}t G\"ottingen, G\"ottingen; Germany.\\
% ...
$^{83}$$^{(a)}$Khalifa University of Science and Technology, Abu Dhabi;$^{(b)}$New York University Abu Dhabi, Abu Dhabi;$^{(c)}$United Arab Emirates University, Al Ain;$^{(d)}$University of Sharjah, Sharjah; United Arab Emirates.\\
% ...
$^{102}$CPPM, Aix-Marseille Universit\'e, CNRS/IN2P3, Marseille; France.\\
% ...
$^{121}$Homer L. Dodge Department of Physics and Astronomy, University of Oklahoma, Norman OK; United States of America.\\
% ...
$^{154}$Raymond and Beverly Sackler School of Physics and Astronomy, Tel Aviv University, Tel Aviv; Israel.\\
% ...
$^{a}$ Also at Affiliated with an institute formerly covered by a cooperation agreement with CERN.\\
$^{b}$ Also at An-Najah National University, Nablus; Palestine.\\
$^{c}$ Also at Borough of Manhattan Community College, City University of New York, New York NY; United States of America.\\
$^{d}$ Also at Center for Interdisciplinary Research and Innovation (CIRI-AUTH), Thessaloniki; Greece.\\
$^{e}$ Also at Centre of Physics of the Universities of Minho and Porto (CF-UM-UP); Portugal.\\
$^{f}$ Also at CERN, Geneva; Switzerland.\\
% ...

\end{flushleft}
